# Helium-assisted, solvent-free electro-activation of 3D printed conductive carbon-polylactide electrodes by pulsed laser ablation

Maciej J. Glowacki[1], Mateusz Cieslik[1], Miroslaw Sawczak[2], Adrian Koterwa[3], Iwona Kaczmarzyk[1], Rafal Jendrzejewski[2], Lukasz Szynkiewicz[1], Tadeusz Ossowski[3], Robert Bogdanowicz[1], Pawel Niedzialkowski[3], Jacek Ryl[1,*]

[1] Gdansk University of Technology, Narutowicza 11/12, 80-233 Gdansk, Poland
[2] The Szewalski Institute of Fluid-Flow Machinery, Polish Academy of Sciences, Fiszera 14, 80-231 Gdansk, Poland
[3] Faculty of Chemistry, University of Gdansk, Wita Stwosza 63, Gdansk 80-308, Poland
[*] Correspondence: Jacek Ryl - jacek.ryl@pg.edu.pl

**Abstract**

The fused deposition modeling is one of the most rapidly developing 3D printing techniques, with numerous applications, also in the field of applied electrochemistry. Here, utilization of conductive polylactic acid (C-PLA) for 3D printouts is the most promising, due to its biodegradability, commercial availability, and ease of processing. To use C-PLA as an electrode material, an activation process must be performed, removing the polymer matrix and uncovering the electroactive filler. The most popular chemical or electrochemical activation routes are done in solvents. In this manuscript, we present a novel, alternative approach towards C-PLA activation with Nd:YAG ($\lambda = 1064$ nm) laser ablation. We present and discuss the activation efficiency based on various laser source operating conditions, and the gas matrix. The XPS, contact angle, and Raman analyses were performed for evaluation of the surface chemistry and to discuss the mechanism of the activation process. The ablation process carried out in the inert gas matrix (helium) delivers a highly electroactive C-PLA electrode surface, while the resultant charge transfer process is hindered when activated in the air. This is due to thermally induced oxide layers formation. The electroanalytical performance of laser-treated C-PLA in He atmosphere was confirmed through caffeine detection, offering detection limits of 0.49 and 0.40 μM (S/N = 3) based on CV and DPV studies, respectively.

## 1. Introduction

The 3D printing technologies found application in diverse branches of industry, such as industrial design, automobiles, architecture, mechanical engineering, biomedical engineering, etc. At present, researchers are exploring new materials, which allows extending the potential application of 3D printing [1–3]. The most popular 3D printing technology is fused deposition modeling (FDM), where thermoplastic materials are used as filaments and the object is printed layer-by-layer. The FDM technology is also the most accessible, due to relatively low costs of purchase and operation, ease of use, and the largest selection of materials for printing, including biodegradable polylactic acid (PLA) and its composites. The high conductivity of PLA composites is most often achieved through the utilization of various electroactive carbon fillers, such as graphene or carbon black (CB). Carbon black (CB) is a very attractive electroactive material due to its high conductivity, biocompatibility, and chemical inertness [4], which main advantage compared to other carbonaceous nanomaterials is its low cost. Due to this fact, CB is the most common electrically conductive additive used in batteries, capacitors, and supercapacitors [5–7] but also printable inks and FDM 3D printable polymers, as used in this case. Possessing higher catalytic activity than highly oriented graphite or carbon nanotubes CB offers fast electron transfer kinetics, a feature particularly important in the case of electrode materials for sensing [8,9].

The accessibility of 3D printing technologies translates to exponentially growing popularity in electrochemical sciences where the technology allows for the easily-accessible manufactory of electrochemical energy storage devices [10–13], and sensing electrodes [14–19], but also electrochemical flow cells [20] or microscopy elements [21]. Here, the application of conductive carbon black-polylactide or graphene-polylactide 3D printed nanocomposite electrodes is flourishing.

The operation of electrochemical sensors is based on the measurement of charge-transfer kinetics alteration at the electrode interface. For 3D C-PLA electrodes to be able to transfer electric signals, their surface must be first activated. The modification procedure goal is to etch the polymer matrix and reveal conductive electroactive carbon filler. The most popular pre-treatment activation procedure for efficient electrode sensitization so far is through immersion in dimethylformamide (DMF) [14,22,23]. There are few recognized attempts to utilize different aprotic solvents, but so far none of them provide comparable surface electroactivity [24,25]. As DMF and numerous other solvents are toxic or carcinogenic there is an urge to limit its use and to search for alternative activation routes. An alternative and possibly more environmentally friendly approach is through anodic and cathodic polarization in neutral electrolytes leading to the formation of small-sized $sp^2$-carbon domains increasing electrode kinetics [26,27]. Particularly promising is also activation carried out under conditions of water electrolysis [28] or enzymatic activation using proteinase K [29].

Laser ablation is a popular method for removing portions of a material from solid surfaces. The laser treatment has not been widely used for processing of PLA-conductive carbon composites, though several such attempts have been reported. Kanczler et al. [30] have used a continuous wave fiber diode laser (wavelength of 970 nm, maximum power of 10 W) for a surface selective laser sintering of PLA powder mixed with microparticles of the carbon black in order to manufacture biocompatible, porous 3D scaffolds promoting adhesion, proliferation, and differentiation of human bone cells. Paula et al. [31] deposited 10 μm thick films composed of the polylactic acid and multilayer graphene fibers on glass substrates using a commercially available filament as a precursor. The resulting layers were illuminated with a $Ti:Al_2O_3$ femtosecond laser (central wavelength of 800 nm, pulses of 50 fs with a repetition rate of 5 MHz) to form electrodes having high-resolution patterns of interdigitated lines. The samples were characterized by an electrical resistivity of $5\times10^3$ Ωm and proved effective as electronic-tongue sensors, able to distinguish substances with different taste properties ($H_2O$, 1 mM NaCl, 1 mM sucrose, 1 mM HCl) without any superposition between them. Ongaro et al. [32] utilized a commercial, $CO_2$ laser cutter Epilog Mini Helix (power, scanning speed, and frequency variable) to engrave microchannels on PLA sheets manufactured by compression molding. Inside the microchannels, electrode tracks were printed with an aqueous graphene ink. PLA monolayers were then bonded together with an adhesive tape to assemble a three-layer hybrid microfluidic device, which was characterized by cyclic voltammetry carried out at 10 mV/s scan rate in the presence and absence of 5 mM potassium ferricyanide, and 5 mM potassium ferrocyanide. The resulting voltammograms displayed clear extremes for ferricyanide/ferrocyanide redox reaction, and no peaks during a negative control.

The goal of this work is to present the capabilities of the laser ablation technique as an alternative route for 3D printed C-PLA electrodes activation. To do so, we present the influence of laser operating parameters and the influence of the gas matrix during ablation on activation efficiency, measured by the increase of the charge transfer process. To the best of our knowledge, this is the first attempt to use laser techniques for localized, selective PLA removal and electrode activation. The approach proposed by us got numerous advantages compared to typically exploited chemical or electrochemical activation methods, such as previously mentioned localized surface modification, allowing structuring the surface and creating patterns, it is rapid and electrolytic-less. Finally, electrochemical caffeine assays based on 3D printed C-PLA electrodes were built and activated with the use of the optimized laser ablation procedure, and the detection limits compared to other carbon-based materials used for caffeine detection, including glassy carbon (GC), graphite, and boron-doped diamond (BDD) [33–35].

## 2. Materials and Methods

The material for 3D-printed electrodes used in this study was commercially available Proto-Pasta Conductive PLA. The given carbon filler content was 20.7 wt.%, which results in 30 Ωcm electric resistivity. It should be noted that the structure of carbon black is analogous to this of graphene, also utilized in commercially-available 3D printing filaments. The basic building unit of carbon black is graphene layers, arranged in parallel to the surface. Since the carbon black particles are spherical, the graphene layers are curved, introducing some heterogeneities in the electronic environment of the carbon atoms at the surface [36]. Moreover, the electrochemical properties of carbon black are on par with these manifested by thermally-reduced graphene oxide, exhibiting a valuable alternative in numerous electrochemical applications, due to carbon black low cost and straightforward production process [37].

Flat electrodes were printed, with dimensions 10x10x2 mm, on Ender 3 Pro 3D Printer (Ender, China). The printing temperature was 200 °C. The electrodes were stored under atmospheric conditions. Electrochemical and physicochemical studies were carried out directly after the printing process. For the cross-section micrographs, a special series of samples with a notch has been printed. Directly after the ablation, those samples were cooled down with liquid nitrogen and then broken in half.

All chemicals were of analytical grade and used without purification. Caffeine, phosphate-buffered saline (PBS), and potassium chloride KCl were purchased from Sigma-Aldrich. All solutions were made from deionized water. PBS was adjusted to pH 7.0 by hydrochloric acid.

The ablation of the 3D-printed C-PLA electrodes was carried out using a pulsed Nd:YAG laser LaserBlast 500 (Quantel, France) operating at 1064 nm and 6 ns pulse duration. The laser is equipped with diffractive optics delivering Top Hat beam profile and square laser spot. Surfaces of the samples have been scanned with a laser spot of 5 × 5 mm. Three different values of an energy density ($I$) – 0.64 or 0.9 or 1.15 Jcm$^{-2}$ – have been tested.

To minimize thermal effects laser pulse repetition was established to 2 Hz. Moreover, the electrodes have been treated with a diverse number of pulses ($n$) – 10, 20, or 30. The atmosphere of the laser ablation was also under examination, with activation carried out in an ambient atmosphere, or an inert gas (helium, purity 5.0). The variability of $I$, $n$, and atmosphere yields 18 combinations of the parameters and enables a thorough study of their impact on the structure and performance of the examined 3D-printed electrode.

Scanning electron microscopy with a variable pressure chamber (VP-SEM) has been selected to observe a topography of the surfaces of the samples and to assess how deep the laser penetrates the electrodes during the ablation. These studies were done with S-3400N microscope (Hitachi, Japan), using 25 kV accelerating voltage and under the pressure of 90 Pa. Before the examination, the electrodes were coated with nanometric layers of gold to increase their conductivity.

The chemical composition of selected electrodes has been examined with X-ray photoelectron spectroscopy (XPS) and Raman spectroscopy. The XPS analyses were carried out using Escalab 250Xi multispectroscope (Thermo Fischer Scientific, USA). The spectroscope is equipped with a monochromatic AlKα source. The high-resolution spectra in *C1s* and *O1s* binding energy range were collected with a pass energy of 15 eV and using an X-ray spot diameter of 250 μm. Charge compensation was controlled through a low-energy electron and $Ar^+$ ions flow by a flood gun. The Raman analysis was performed using a Raman microscope (InVia, UK). Spectra were recorded in a range of 120–3200 $cm^{-1}$ using an argon-ion laser emitting at 514 nm and operating at 1% of its total power (50 mW) to avoid PLA melting. Data were smoothed (with a Savitzky-Golay method, 15 points, 2nd polynomial order). Each sample was analyzed in five randomly selected points on its surface. The spectrum of every single point at the sample was recorded as an accumulation of five scans. The exposition time of the single scan was set to 20 s. For the Raman analysis of sample cross-section, PLA plates were cooled down in liquid nitrogen and mechanically broken. On the sample cross-section, Raman spectra were recorded in the middle of sample thickness and near the surface irradiated with a laser, using the x100 microscope objective. The contact angle measurements were conducted using Drop Shape Analyser (DSA) 100, (Krüss, Germany). All analyses were performed at room temperature, using 2 μm deionized water drop volume. The contact angles were determined using the sessile drop method. The analysis of the shape of the drop was performed using the Young Laplace fitting method. All measurements were repeated ten times.

Every activated electrode was subjected to electrochemical analysis to verify the kinetics of the charge-transfer process due to activation. Electrochemical impedance spectroscopy (EIS) and cyclic voltammetry (CV), and differential pulse voltammetry (DPV) studies were performed on Autolab 302N potentiostat (Metrohm, Switzerland), controlled by NOVA 2.4 software. The electrochemical studies were carried out in a three-electrode setup, with C-PLA as the working electrode, Ag|AgCl (3 M KCl) as the reference electrode, and Pt wire as the counter electrode after initial conditioning for 10 minutes. Studies were done in 5 mL electrochemical cells. The EIS activation efficiency tests were done in 0.1 M KCl, and in the 10 kHz – 0.1 Hz frequency range, with 15 perturbation signals per frequency decade. The amplitude of the perturbation signal was 10 mV and the measurements were carried out in potentiostatic mode, in open circuit potential conditions. The detection of caffeine was performed in 0.01 M PBS solution (pH = 7.0) in increasing concentration of caffeine from 0.01 mM to 1 mM. The CV and DPV procedures for caffeine detection were carried out in the polarization range from +0.5 to +1.6 V vs Ag|AgCl.

## 3. Results and discussion

### 3.1. The influence of laser energy density, number of pulses, and ablation gas atmosphere

The efficient 3D printed C-PLA composite electrode activation process is based upon the effective removal of the surface polymer matrix, which would uncover the electroactive carbon filler, avoiding the oxidation processes and degradation of the polymer matrix, leading to sub-optimal electrode characteristics. The process may be optimized through careful consideration of the internal factors affecting laser ablation, to be carried out as the first step of this study. Different laser energy densities ($I$) and the number of short irradiation cycles ($n$) were tested. The effect of the proposed activation conditions was verified using EIS analysis, comparing the results with the C-PLA electrode before the activation.

The Bode plots in **Fig. 1** shows the impedance modulus $|Z|$ vs. applied perturbation frequency for each studied C-PLA electrode, with the corresponding phase angle variation ($\theta$) shown in the inset. When comparing different laser energy densities, the smallest impedance modulus was observed at $I = 0.64$ Jcm$^{-2}$ and rises with the energy density increase. This is best noticed at the lowest frequency range, where the $|Z|$ value represents the value of the charge transfer resistance through the electrode interface. Regardless of the applied laser operating parameters, the activation process leads to the decrease in charge transfer resistance by not less than two orders of magnitude, down from approx. 4.20 MΩ, recorded for the inactive C-PLA electrode. The impedance modulus recorded at $f$ = 0.1 Hz after laser ablation treatment was equal to 2.27, 4.32, and 8.90 kΩcm$^2$, for 0.64, 0.90, and 1.15 Jcm$^2$ laser energy densities, respectively. The analysis of the phase angle spectrum reported in the inset of **Fig. 1a** shows that the impedance spectra of each surface-modified C-PLA electrodes are characterized with a quite similar value of the time-constant, evidenced by a small shift towards lower frequencies range when comparing between the most efficient activation at 0.64 Jcm$^{-2}$ and different laser energy densities. This outcome suggests a similar mechanism in the charge transfer process observed at each of the activated C-PLA electrodes.

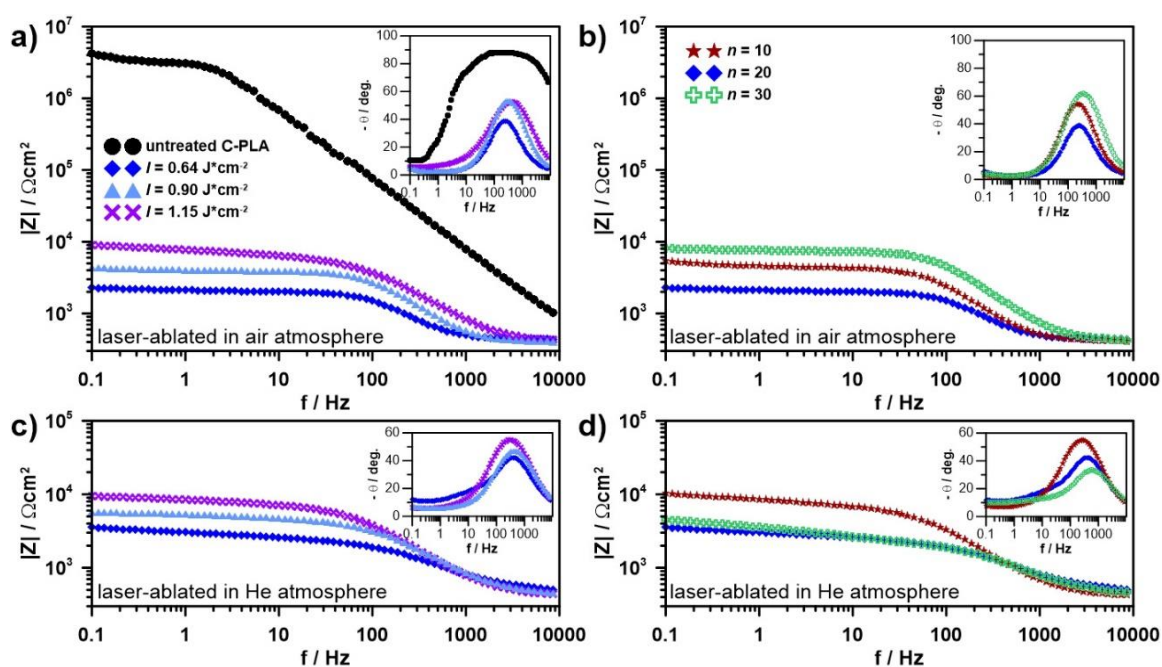

**Figure 1**. The impedance Bode plots (phase shift plots in the inset) registered before and after laser-induced activation of C-PLA electrodes a,b) in air, c,d) in He atmosphere; studies carried out after ablation with various a,c) laser energy density $I$ ($n$ = 20), b,d) numbers of short laser pulses $n$ ($I$ = 0.64 Jcm$^{-2}$). Studies were carried out in 0.1 M KCl, after 10 minutes of electrode conditioning.

The number of laser pulses appears to have some effect on the C-PLA electroactivity, while the most prolonged ablation duration consequences in the highest recorded impedance modulus (and charge transfer resistance), equal to 7.99 kΩcm$^2$. On the other hand, too short of an irradiation cycle consequence in partial surface area activation, negatively affecting the resultant electroactivity.

In order to evaluate the influence of the PLA oxidation on the laser-activated C-PLA electrode charge transfer kinetics, we have performed the same procedure in the noble gas matrix (helium). Interestingly, the inert gas atmosphere had a negligible influence on the electrochemical activity after laser ablation treatment, thus suggesting that surface oxidation is not the reason behind hindered electroactivity at higher energy densities (see **Fig. 1c**).

One crucial difference emerges comparing the cross-section topography of C-PLA electrodes treated with various laser energy densities, which is the roughness of laser-etched crater edges, seen in the insets of **Figs. 2c and 2d**. The kinetic energy of the molecules in plasma translates on the temperature-induced vibrations and rotations within the ablated PLA, whose nature is more local at low energy densities. At low energy densities, the ablation is characterized by the more local character of material degradation and ragging the edges of laser-etched craters. The higher the energy density, the larger the areas of elevated temperature under laser interaction. When a laser with certain energy is irradiated on the C-PLA surface it becomes molten, breaking the polymer chains under the action of the re-melting mechanism, and making the surface material to flow to unfilled spots. A similar mechanism was described in detail in the studies dedicated to PLA laser polishing [38,39]. Based on the above-presented mechanism and SEM cross-sections we hypothesize that re-covering of active carbon filler taking place at higher laser energy densities is restraining the electrochemically active surface area. The hypothesis is in good agreement with similar EIS findings obtained in air and He atmosphere.

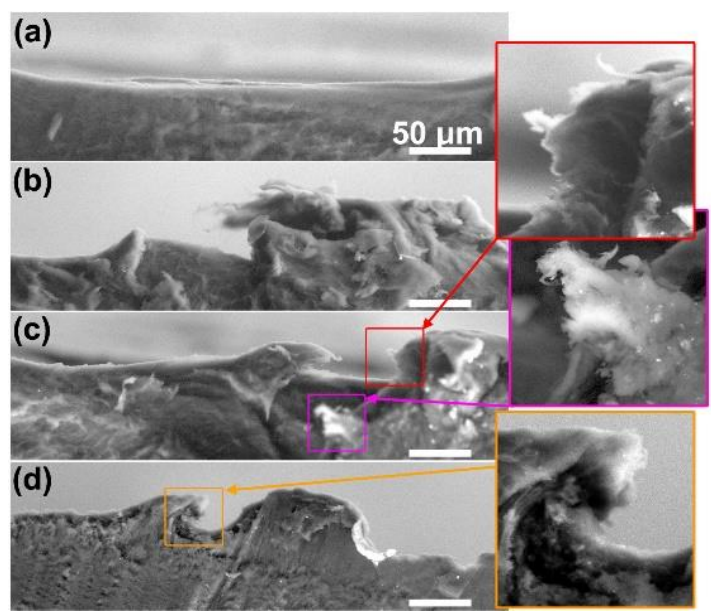

**Figure 2.** Cross-section of C-PLA electrodes: a) untreated for reference, b-d) subjected to laser ablation in air atmosphere (n = 20) at various energy densities I: b) 0.64 Jcm$^{-2}$, c) 0.90 Jcm$^{-2}$, d) 1.15 Jcm$^{-2}$.

The VP-SEM micrographs presented in **Fig. 2** also reveal that 20 short laser pulses are enough to penetrate the C-PLA material approximately 70-100 μm underneath the surface and that the penetration depth does not depend greatly on laser energy density. The above result allows us to conclude that the C-PLA electrode roughness is similar in each studied case and does not have the key influence when it comes to the development of the electrochemically active surface area (EASA).

On the other hand, carrying out the ablation process in the He atmosphere has a visible effect on the C-PLA electroactivity when considering the influence of the number of laser pulses. Here, unlike in the case of air-assisted irradiation, the detrimental effect of the prolonged laser ablation is significantly lower, as testified by the impedance results in **Fig. 1d**. These findings suggest degradation of the electroactive carbon species with ablation time and in the presence of oxygen, as will be further discussed based on Raman spectroscopy analyses.

The micrographs in **Fig. 3** allows us to compare the activated C-PLA electrode topography and depth of penetration depending on the gas atmosphere of the ablation process. The comparison was performed in the optimized activation process operating conditions ($I = 0.64$ Jcm$^{-2}$, $n = 20$). When studying topography at low magnifications (x100, **Fig. 3a**) one can spot the remnants of the partially etched polymer matrix, seen as the bright colors due to its non-conductive nature. The gas atmosphere used for the activation process does not seem to have a significant influence on the topography of the C-PLA electrode. **Figs 3b** and **3c** present the micrographs in the cross-section, revealing a similar penetration depth, in both cases reaching 30-40 μm. Furthermore, the high magnification micrographs of the activated C-PLA surface show highly similar effects of the polymer matrix removal (**Figs 3d,e**). Here, the light-green color highlights the carbon black agglomerates, uncovered by the laser ablation treatment [40]. It is evenly distributed at the composite surface. Its local amounts at certain electrode areas may vary, regardless of the studied environment.

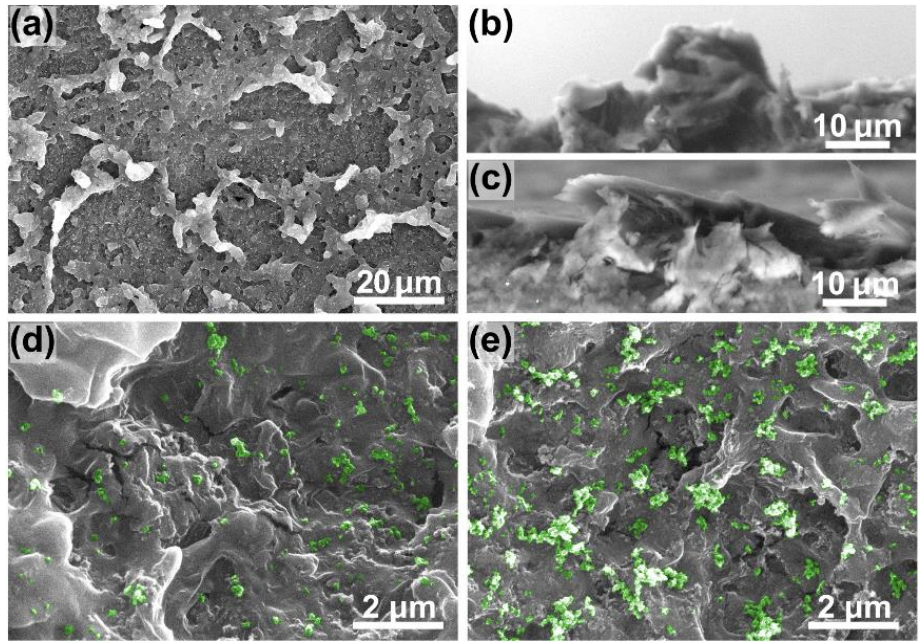

**Figure 3.** Micrographs showing a,d,e) topography and b,c) cross-section of C-PLA samples ablated in different atmospheres: a,b,d) air atmosphere; c,e) He. Ablation parameters: $n = 20$, $I = 0.64$ Jcm$^{-2}$.

The reference Raman spectra were recorded on the cross-section of the laser-irradiated electrode (parameters of ablation: $n = 20$, $I = 1.15$ Jcm$^{-2}$), in an area completely unaffected by the laser. This region of the electrode has been considered a core of the sample, an area composed of C-PLA which was melted during the printing process, but immediately thereafter covered with successive layers of the material, greatly limiting its exposure to the ambient air during the cooling. Next, the Raman spectra were recorded for a surface of the untreated electrode, as well as for electrodes processed with a laser using an energy density of 1.15 Jcm$^{-2}$ and a dose of 20 laser pulses per electrode surface (1 cm$^2$), in air and helium atmosphere. The Raman spectra are dominated with D and G bands centered at 1361 and 1596 cm$^{-1}$ respectively and derived from carbon black filling [40,41], as seen in **Fig. 4a**. The D band is associated with disordered sp$^3$-hybridized carbon featured as defects or impurities in the carbon materials, the G band is associated with the E$_{2g}$ phonon modes of the sp$^2$-bonded carbon, and a broad 2D peak around 2722 cm$^{-1}$ is the second order of zone-boundary phonons [36,42,43]. A (D + G) band at 2957 cm$^{-1}$ was due to the defects in the sp$^2$ sites. The I2D/IG ratio changes for samples processed with the laser. For pristine material, the value of the I2D/IG ratio is about 0.12 and increases to 0.14 and 0.20 for samples processed with laser in air and helium atmosphere respectively.

The most visible effect of thermal processing of C-PLA is the increase of background photoluminescence. In the core of the sample, the PLA material is homogenous, which is confirmed by a negligible value of the standard deviation. Additionally, no differences were observed between spectra recorded in the middle of the sample's cross-section and the proximity of the surface. It should be noted that the surface of the sample, even if unprocessed by the laser, shows increased background fluorescence in comparison to the polymer analyzed in the core. This effect can be explained with PLA oxidation processes occurring on the surface as a result of contact of the molten material with air. The level of the background photoluminescence increases further after laser processing in the air. This phenomenon, which originates from structural defects caused by thermal quenching and re-solidification [44,45], indicates degradation of PLA during laser ablation in the presence of oxygen. The opposite effect is observed in the case of laser ablation in the helium atmosphere when the background photoluminescence decreases compared to the level recorded for the untreated surface. This can be explained by laser removal of the top layer of the material. At the same time, degradation processes during ablation in the oxygen-free atmosphere occur less intensively.

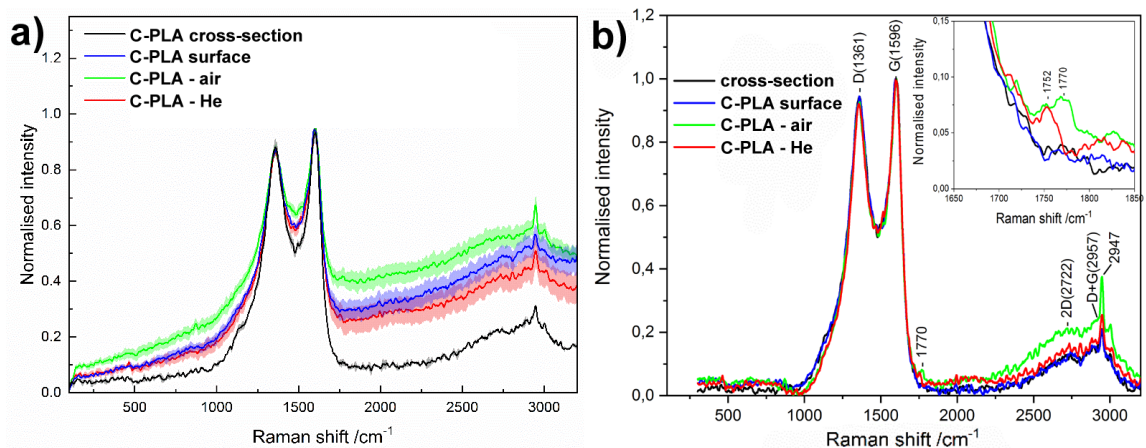

**Figure 4**. a) Raman spectra recorded for: printed C-PLA electrode cross-section, the unprocessed surface of the 3D printed electrode, and electrode surface after laser treatment in air and helium atmosphere. The pattern indicates the standard deviation of spectra measured in different points; b) Raman spectra after removing background photoluminescence, with the zoomed-in 1650 to 1850 cm$^{-1}$ spectral range in the inset.

Removing background photoluminescence, as seen in **Fig. 4b**, allows exposing weak Raman bands of the polymer. The Raman spectra of laser processed C-PLA expose weak bands in the region from 1750 to 1776 cm$^{-1}$ (**Fig. 4b** – inset). According to literature, bands at 1750, 1766, 1770, and 1776 cm$^{-1}$ can be assigned to C=O stretch vibrations. Additionally, the band located at 1770 cm$^{-1}$ is characteristic of the amorphous phase of PLA polymer and its intensity can be used to determine the crystallinity of the material [46]. In our experiment evident increase of 1770 cm$^{-1}$ band is observed for samples irradiated with laser in ambient air atmosphere. For samples irradiated in He atmosphere the Raman signal at 1770 cm$^{-1}$ is much weaker. In general, the appearance of the amorphous phase after laser irradiation can be explained by a high cooling rate due to the large temperature gradient induced by laser ablation. It can be supposed that increased intensity of the band at 1770 cm$^{-1}$ in the case of samples irradiated in the presence of oxygen in the ambient air may be associated with an increased number of defects in PLA structure caused by oxidation processes. In the spectral range of 2800 – 3000 cm$^{-1}$ the CH$_3$ symmetric (2881, 2947 cm$^{-1}$) and asymmetric (3000 cm$^{-1}$) bands can be observed. The increased intensity of the band centered at 2947 cm$^{-1}$ correlate with the decrease of PLA crystallinity [46].

Next, the X-ray photoelectron spectroscopy was used to evaluate the surface chemistry modification as a result of C-PLA irradiation in air and helium atmosphere. These studies were compared with surface chemistry of the reference freshly 3D printed C-PLA electrode and collectively presented in **Fig. 5**.

The high-resolution XPS analysis in the C1s energy range revealed a complex multicomponent spectrum, with not less than four different spectral components, partially overlapping each other (**Fig. 5a**). Taking into consideration PLA chemistry, three different carbon chemical states are to be expected, namely C-C, C-O, and C=O, and the expected ratio is 1:1:1. These three chemical states are represented by *C2*, *C3*, and *C4* peaks, respectively. It can be noticed, that in contrast to the theoretical 1:1:1 ratio expected for pure untreated PLA, the *C2* C-C/C-H peak is significantly higher, indicating the excess of hydrocarbons on untreated PLA surface, to be explained by the contribution of adsorbed adventitious hydrocarbons contamination from the atmosphere [47,48]. The primary component from sp$^2$-carbon within the conductive carbon black filler (so-called graphite peak) is expected to appear negatively shifted versus the aliphatic hydrocarbons component [36,40,49,50]. Indeed, the *C1* peak, reflecting the share of surface carbon black filler is shifted at -1.1 eV vs *C2* peak. Its share in the reference sample is on part with the *C2* component, demonstrating its significant amount in the most outer part of the electrode surface (XPS depth analysis is approx. 5 nm).

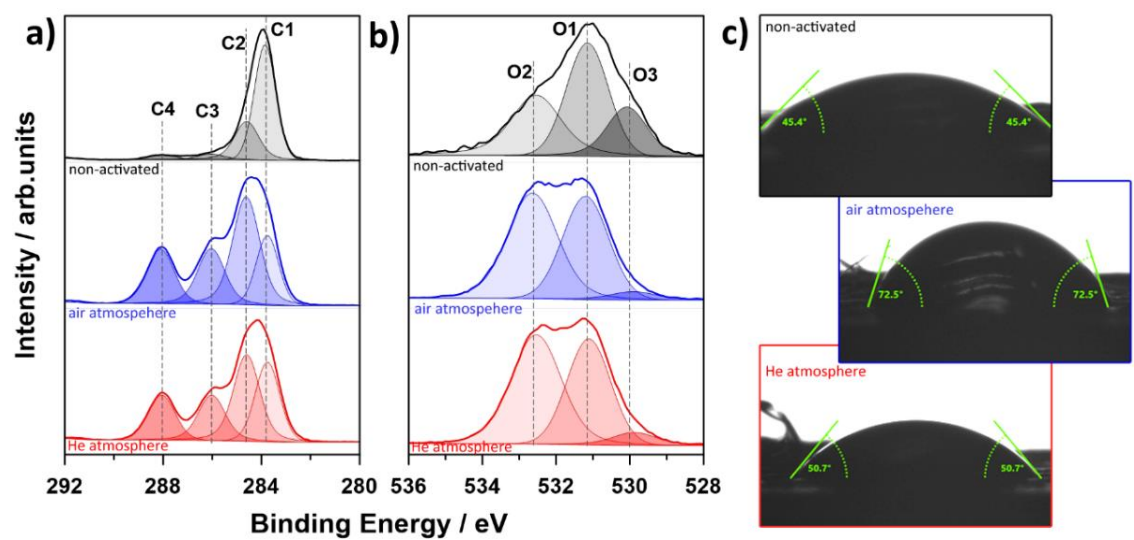

**Figure 5.** Surface carbon and oxygen chemistry of 3D printed reference C-PLA before modifications, and C-PLA electrodes after laser ablation in air atmosphere, and He atmosphere; a) C1s and b) O1s high-resolution XPS spectra. c) contact angles measured for the corresponding C-PLA electrodes.

The O1s spectra recorded for the reference C-PLA electrode were also deconvoluted using four spectral components (**Fig. 5b**). The two dominant ones *O1* and *O2*, located at binding energies of 531.2 and 532.6 eV, should be ascribed to C=O and C-O bonds, respectively [48,51]. The *O1*:*O2* ratio is 0.8:1, which is lower than the theoretical 1:1 ratio expected of PLA. Again, the higher share of C=O bonds is to be explained with adventitious carbon contamination. The *O3* is problematic in unequivocal identification and may hypothetically represent some metal-oxygen bonds in the impurities introduced by the C-PLA filament manufacturer [52]. Finally, the last component is only present in the spectra recorded for the reference C-PLA and is most likely connected with surface-chemisorbed water molecules resultant from prolonged air exposure of 3D printed electrodes [51,53]. The details of the XPS analysis are summarized in **Table 1.**

**Table 1.** The surface chemical composition (in at. %) of the 3D printed C-PLA electrode, as well as after laser ablation in air or helium atmosphere, based on the high-resolution XPS analysis.

| | Chemical state | BE / eV | Reference | Laser ablated in air | Laser ablated in helium |
|---|---|---|---|---|---|
| C1s | C1 | 283.7 | 22.8 | 16.9 | 27.5 |
| | C2 | 284.8 | 22.2 | 28.2 | 23.9 |
| | C3 | 286.0 | 15.9 | 13.8 | 11.9 |
| | C4 | 288.1 | 14.5 | 12.7 | 11.5 |
| O1s | O1 | 531.2 | 9.1 | 12.3 | 11.7 |
| | O2 | 532.6 | 11.8 | 14.4 | 11.6 |
| | O3 | 529.9 | 1.0 | 1.8 | 1.9 |
| | O4 | 534.1 | 2.6 | -- | -- |

When compared the C-PLA electrode pre-treated by laser ablation in the air or helium atmosphere with reference C-PLA one can see a few key differences in the contribution of individual components. Interestingly, the contribution of electroactive carbon nanoparticles, represented by *C1* peak decreases for ablation in the air atmosphere, which is assisted with the growth of *C2* peak contribution. The *C2*:*C3*:*C4* ratio changes from 1.5:1.1:1.0 for the reference sample up to 2.2:1.1:1.0, which might suggest the increasing share of hydrocarbon contaminants. Simultaneously, the amount of the PLA signal at the electrode surface is slightly decreasing, which can be directly tracked through C-O and C=O C1s peak contribution. The ablation in the air is assisted with higher surface oxygen concentration, adsorbed, or chemically bonded, as can be seen in **Table 1**.

However, a different explanation for significantly higher *C2* peak contribution is more plausible and supported with both literature and Raman spectroscopy data. The filler may be partially oxidized as a result of the laser treatment, with a consequence in a shift of the C1s peak towards higher binding energies [36,54]. Given the correctness of the hypothesis, the formation of oxidized carbon black would explain a significantly decreased amount of *C1*:*C2* ratio (from 1.0:1.0 for the reference sample, down to 1.0:1.7 for C-PLA sample after ablation in air atmosphere).

Further evidence is provided when analyzing the surface chemistry of C-PLA after irradiation in the helium atmosphere. First and foremost, the share of the *C1* spectral component is the most prominent for this sample, among all studied C-PLA electrodes (27.5% vs 22.8 for reference C-PLA and 16.9 for C-PLA ablated in air atmosphere). At the same time, different studies on pure PLA reveal the C1s peak ratio to be significantly closer to theoretical 1:1:1 PLA if laser-treated in He atmosphere [48]. Notably, the surface PLA removal and activation seem to be even more efficient in our case, as testified by a further decrease of *C3* and *C4* components. Finally, the contribution of the *C2* peak is similar to the reference sample.

The measurement of the contact angle is a very useful tool, providing information about changes in the polarity of target surfaces, before as well as after their modification [55,56], therefore, we have decided to carry out contact angles studies for the C-PLA electrodes. The results of this analysis reveal the wettability dependence on the preparation method (**Fig. 5c**). In all cases, the value of the obtained contact angle does not exceed 90°, indicating that all investigated C-PLA surfaces are hydrophilic [57]. The contact angle determined for the reference C-PLA electrode before laser ablation treatment was 45.4°, which is slightly lower from values reported in the literature due to PLA remelting during the 3D printing operation [58–60]. Next, after laser ablation in the air atmosphere, the value of the contact angle increases to 72.5°, which directly indicates that the ablation in the air atmosphere causes a decrease in surface hydrophilicity. On the other hand, changes in hydrophilicity occurring at the C-PLA surface after ablation in He atmosphere are not so significant in comparison to reference C-PLA and the contact angle reaches 50.7°. The contact angle studies follow the XPS results, testifying minor C-PLA surface chemistry modification as a result of laser ablation in both air and He-based atmosphere.

Lu et al. [61] reported that carbon black exhibits a water contact angle of 89°, while the oxidized carbon black displays hydrophilic contact angles of approx. 60° [62], due to the incorporation of carbonyl and carboxyl polar groups. Furthermore, the contact angle of PLA/CB composites was reported to achieve *ca.* 45° thanks to decay and dissipation of static electricity of pristine PLA having a contact angle of approx. 83° [63].

We have observed that the air-based laser ablation results in a stronger hydrophobic surface than treatment in the noble gas. This fact suggests that more pristine carbon black is exposed by air-based treatment, while this finding is not fully supported by XPS. Wetting interactions reveal complex nature here thanks to the hybrid composite structure inducing heterogeneous distribution of surface free energy. This led us to the conclusion, that contact angles studies of the ablation process could be applied only for qualitative studies supporting data extracted by XPS or Raman due to developed surface morphology along with complex structural composition.

**3.2. The evaluation of laser-activated C-PLA electrode for caffeine detection**

The mechanism of caffeine electrooxidation using electroanalytical methods is presented in (**Figs 6a,b**). This mechanism includes four electrons and four protons and consists of two steps. The first step is slow and leads to the formation of the substituted derivative of uric acid, while the second step is fast and causes further oxidation and formation of uric acid 4,5-diol analog that undergoes further fragmentation as a consequence of oxidation [64,65]. Thus, the detection of caffeine present in an analyte solution may be easily performed by measuring its electrochemical response. Two electrochemical techniques were used for this task, namely CV and DPV, and the obtained results are presented in **Figs 6c** and **6d**, respectively. This test was performed in 0.01

M PBS solution (pH = 7.0) containing 1 mM of caffeine. The peak potentials of caffeine oxidation appear at 1.5 V and 1.4 V vs Ag|AgCl for CV and DPV respectively, corroborating other findings [66–69].

The caffeine oxidation peaks are best developed at the C-PLA electrode, when activated in He atmosphere, and reach over 200 µA, in comparison to the remaining studied samples. The signal strength is incomparably smaller for the C-PLA after ablation in the air atmosphere, the reference untreated C-PLA electrode remained electrochemically inactive. A similar conclusion may be drawn from the DPV experiment. The electrochemical activity of the C-PLA electrode after ablation in He can be explained with effective PLA polymer layer removal by the laser beam, thus exposing the electroactive carbon to the highest extent. One should take into consideration, that factors such as electrode contamination with caffeine oxidation products, or operating near the polarization range reported for the electrochemical C-PLA activation [27,70] may have some influence on the oxidation kinetics during the consecutive polarization cycles.

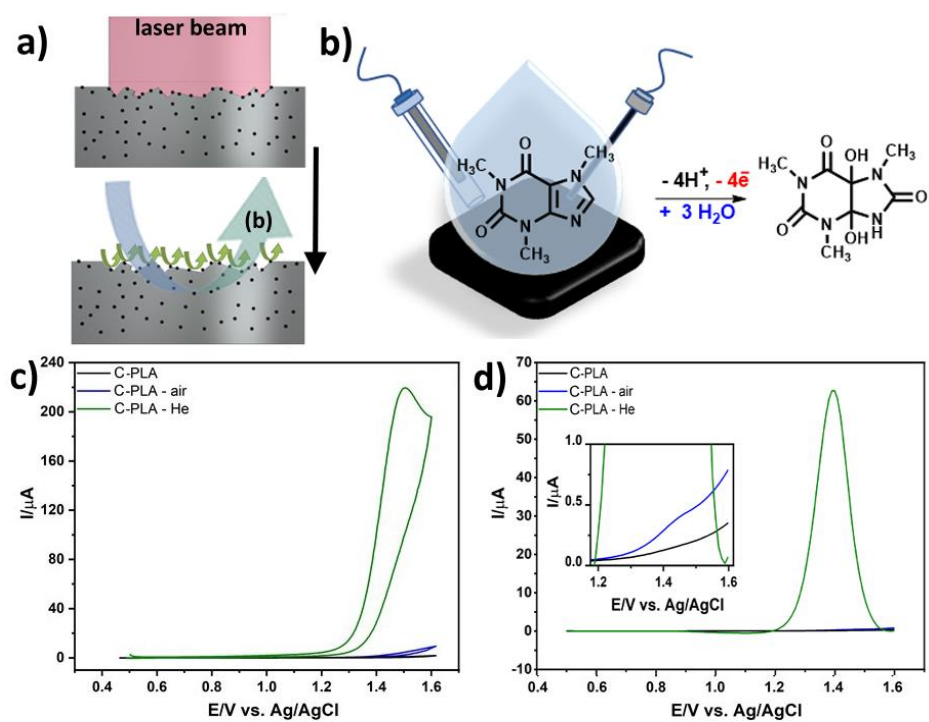

**Figure 6.** a,b) Mechanism of electrochemical caffeine oxidation at the C-PLA electrode, c) CV and d) DPV scans for caffeine detection at different C-PLA electrodes. Measurements for 1 mM caffeine concentration in 0.01 M PBS, before and after C-PLA electrode activation in air or helium atmosphere. Laser ablation parameters: $n = 20$, $I = 0.64$ Jcm$^{-2}$.

The above-presented experiment allowed us to verify the highest electroactivity of the C-PLA surface activated in He atmosphere. The quantitative determination of caffeine was performed in the caffeine concentration ($C_{caffeine}$) range from 0.01 mM to 1 mM. The CV measurements of caffeine oxidation indicate that the peak current ($I_{caffeine}$) in the obtained voltammograms increases linearly with the increasing caffeine concentration (**Fig. 7a**). The relationship between caffeine oxidation current and caffeine concentration is presented in the inset (**Fig. 7a** inset). The C-PLA electrode after ablation in He expresses a good linear response with the linear regression given by eq. (1):

$$I_{caffeine} \text{ (μA)} = 0.2110 \, [C_{caffeine} \text{ (M)}] + 6.727 \quad (1)$$

and with correlation coefficient ($R^2 = 0.990$). The limit of detection (LOD) was calculated to be 0.49 μM (with the S/N ratio of 3). The obtained detection limit is comparable with LOD obtained for modified glassy-carbon electrodes by CV [71] or square-wave voltammetry [72], which we found to be very satisfactory.

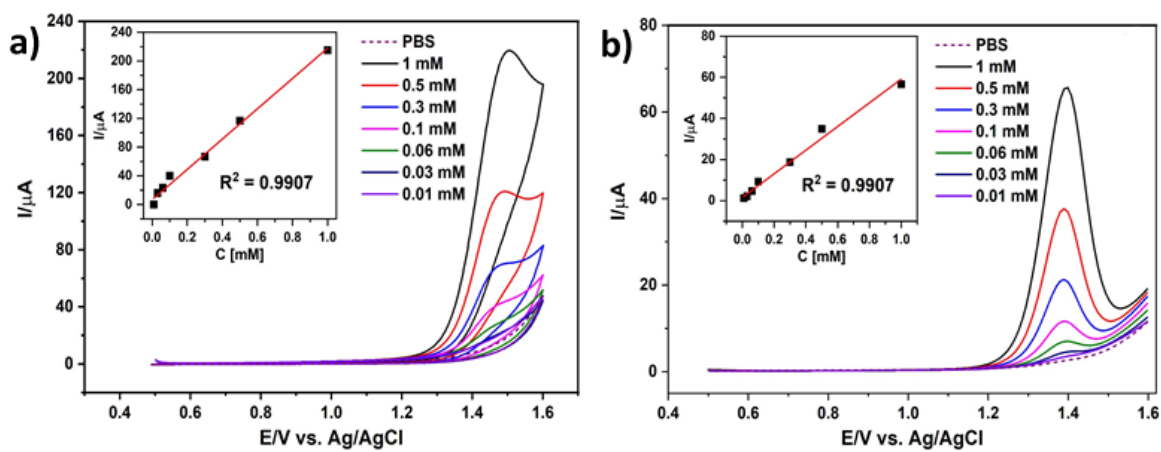

**Figure 7.** CV a) and DPV b) scans for caffeine detection at the C-PLA electrode activated in helium atmosphere. Laser ablation parameters: $n = 20$, $I = 0.64$ Jcm$^{-2}$. Electrolyte 0.01 M PBS.

The DPV experiment provides a similar result as the CV studies, the caffeine determination by DPV is shown in **Fig. 7b**. The relationship between caffeine oxidation current and caffeine concentration is presented in **Fig. 7b** inset. The linear response was observed also in this case, with the regression equation given by eq. (2):

$$I_{caffeine} \text{ (μA)} = 0.0636 \, [C_{caffeine} \text{ (M)}] + 2.288 \quad (2)$$

and with correlation coefficient ($R^2$ = 0.997). The LOD was calculated to be 0.40 μM (S/N = 3), which is very good compared to other studies using DPV [33,73,74]. The obtained CV and DPV results allow us to confirm high efficiency and good reproducibility of quantitative evaluation of the caffeine presence with the use of C-PLA electrodes after surface activation by laser ablation in He atmosphere.

**4. Conclusions**

In summary, investigations focused on the laser treatment of Proto-Pasta Conductive PLA electrodes revealed high efficiency of the ablation process enabling electro-activation of their surfaces.

The laser-induced activation process involves the effective removal of polymer matrix from the 3D printed C-PLA composite electrode surface suppressing the electroactive carbon black oxidation and degradation of the polymer matrix. Microimaging exhibited that just 20 of short laser pulses were enough to achieve the above-mentioned effect penetrating electrode to approx. 100 μm in-depth.

It was found that the penetration depth does not depend on the laser energy density, while much more on the treatment repetition and process atmosphere. The main role during surface treatment plays a physical effect of C-PLA surface melting and breaking the polymer chains during re-melting processes induced by laser irradiation.

The laser ablation procedure was conducted in the inert gas matrix (helium) to avoid the C-PLA oxidation process in the air, which is hindering the charge transfer through the electrode interface. The achieved both contact angles and XPS data showed the minor influence of the laser treatment in a noble gas atmosphere on the modification of C-PLA chemistry, while delivering a highly electroactive surface. The undesirable effects like surface oxidation and oxidized carbon black phases formation at the C-PLA electrode surface were achieved during the air-assisted ablation process similar to the thermal treatment approach.

The electroanalytical performance of laser-treated C-PLA electrodes was proved through the caffeine detection process involving CV and DPV techniques. The caffeine oxidation peaks were considerably enhanced at the C-PLA electrode, activated in the helium atmosphere in contradiction to the air-assisted laser activation or freshly printed untreated surfaces.

## 5. Acknowledgements

This work was supported by The National Centre for Research and Development Techmatstrateg 347324/12/NCBR/2017.